\documentclass[
reprint,
superscriptaddress,
amsmath,amssymb,
aps,
pra,
longbibliography,
]{revtex4-1}
\usepackage{float}
\usepackage{xcolor}

\usepackage{graphicx}
\usepackage{dcolumn}
\usepackage{bm}
\usepackage[colorlinks=true, pdfstartview=FitV, linkcolor=blue, citecolor=blue, urlcolor=blue]{hyperref}
\usepackage{dsfont}

\usepackage{datetime}

\usepackage[separate-uncertainty=true]{siunitx}
\usepackage[siunitx]{circuitikz}

\graphicspath{{figures/}}

\begin{document}

\preprint{APS/123-QED}

\title{The Stochastic Guitar}

\author{Andreas Eggenberger}
\affiliation{Department of Physics, ETH Zurich, Otto-Stern-Weg 1, 8093 Zurich, Switzerland}
\author{Alexander Eichler}
\affiliation{Laboratory for Solid State Physics, ETH Z\"{u}rich, CH-8093 Z\"urich, Switzerland.}
\affiliation{Quantum Center, ETH Zurich, CH-8093 Zurich, Switzerland}

\date{\today ~ \currenttime}

\begin{abstract}
Stochastic physics is a central pillar of modern research in many fields, but is rarely presented to undergrad students in a hands-on experiment. Here, we demonstrate how a human-scale, simple, and affordable experimental setup can be used to fill this gap, and to illustrate many advanced concepts in a step-by-step approach. Based on a metal wire (such as a guitar string), our setup facilitates the observation of fluctuating dynamics in the time domain, the frequency spectrum, and in the rotating phase space. The latter allows introducing time-dependent cross-correlations between the sine and cosine quadratures of the stochastic motion, which feature deterministic order even in the absence of any deterministic forces.
\end{abstract}

\maketitle

\section{Introduction}
The driven and damped harmonic oscillator is a textbook model equation that every physics student is familiar with. Application of this model can be found in every field of physics, from quantum optics to particle physics, biophysics, and astronomy. For that reason, the harmonic oscillator is also a popular experiment on display in many student labs and practical lecture experiments. Students typically learn about the oscillator's resonance frequency $\omega_0 = 2\pi f_0$, the rate of energy loss to the environment $\Gamma$, and the corresponding quality factor $Q=\omega_0/\Gamma$. In the vast majority of student experiments, the harmonic oscillator is treated as a deterministic system without any noise.

The physics of the harmonic oscillator becomes much richer when a stochastic driving force is added. In this case, the oscillation has no definite amplitude $X$ and phase $\phi$, but explores different values thereof as a function of time in a stochastic random walk. It is therefore generally impossible to predict $X$ and $\phi$ at any point in time in the future, and only statistical probabilities can be calculated. Such a scenario is immensely difficult to grasp for many undergraduate students for several reasons: (i)~Dealing with probabilities instead of deterministic equations of motion requires very abstract thinking. (ii)~Stochastic physics is not common, or at least not obvious, in our everyday lives. The systems that offer themselves for such studies, such as nanomechanical resonators, usually operate on scales of amplitudes and frequencies that are far from our intuition and hard to measure. (iii)~Due to the stringent requirements for observing e.g. the Brownian motion of molecules, student labs rarely offer experiments to explore stochastic physics. Most students therefore only come into contact with this important aspect of physics when they perform projects in research groups. This is deplorable, as stochastic physics is not only a crucial element in many contemporary fields of physics, but also offers an excellent preparation for the stochastic phenomena encountered in quantum mechanics.

In this work, we describe a student lab experiment that allows observing the harmonic oscillator with a stochastic driving force. The experiment is based on a metal string whose fundamental resonance frequency is around \SI{30}{\hertz} and whose oscillation can be visible to the naked eye. In spite of these ``human'' scales, the string can be driven in a stochastic manner, and its displacement can be measured digitally with high precision. This results in rich measurement data and complex concepts, such as the rotating phase space, the power spectral density of different types of noise, the Parseval theorem, and the equivalent of under- or overdamped oscillations around the stationary solutions of an out-of-equilibrium system. The experiment offers motivated students a hands-on approach to many fascinating concepts, and a bridge to advanced experiments in research labs. In addition, the experiment is ideally suited for public demonstrations thanks to its accessible character: everybody loves a guitar!

\section{Experimental Setup}
\begin{figure}[h!]
    \includegraphics[width=\columnwidth]{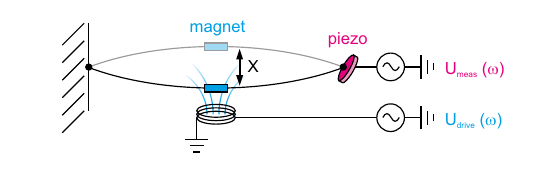}
    \caption{Sketch of the setup}
    \label{fig:Fig1}
\end{figure}

The heart of our experiment is a metal string with a cross section of $0.15 \times\SI{0.9}{\milli\meter}$  made out of stainless steel, as inspired by previous experiments~\cite{Leuch_2016,eichler2018parametric}. The string is suspended between two clamping points with a separation of \SI{275}{\milli\meter}, see Fig.~\ref{fig:Fig1}. The tension applied to the string can be used to tune its resonance frequencies. For our demonstration, we selected a relatively low tension, which results in low resonance frequencies and low frequency drift in response to temperature changes in the room. In the center of the string, we place a neodymium magnet. This allows us to apply a force to the string by running a current through a coil below the magnet. The displacement of the string is recorded through a piezoelectric crystal embedded in one of the clamping points. In this crystal, the pressure that the string exerts onto the clamping point is transduced into a voltage that we measure with a Zurich Instruments MFLI lock-in amplifier (any analog-digital converter would do).

\begin{figure}[t]
        \includegraphics[width=\columnwidth]{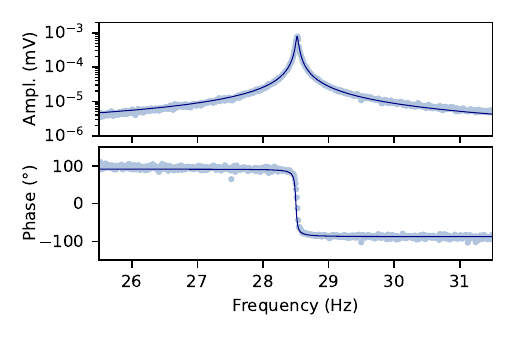}
        \caption{Amplitude and phase of a measured frequency sweep (light blue) plotted versus the results of Eq.~\eqref{eq:response_X} and \eqref{eq:response_phi} (dark blue), using $f_0 = \omega_0/2\pi =\SI{28.52}{Hz}$, $\Gamma =  \SI{0.209}{\hertz}$, and a global phase offset of \SI{50}{\degree} as free parameters. The measurement used \SI{15}{second} waiting time between each point and steps of \SI{10}{\milli\hertz}.
        }
        \label{fig:Fig2}
\end{figure}

\section{Deterministic Response}
To probe the basic properties of our string, we apply an oscillating current to the coil. The current generates a force $F(t) = F_0 \cos(\omega t + \theta)$ with amplitude $F_0$ and frequency $f = \omega/2\pi$. For fixed forcing parameters, the string ``rings up'' until it responds with stable oscillations at the driving frequency. This behaviour can be understood from the well-know equation of motion of a single harmonic resonator
\begin{align}\label{eq:EOM}
    \ddot{x} + \omega_0^2 x + \Gamma \dot{x} = \frac{F(t)}{m}\,,
\end{align}
where $x$ is the displacement of the string and $m$ is its effective mass. With the Ansatz $x(\omega) = X(\omega)e^{i(\omega t +\phi(\omega))}$, we can solve Eq.~\eqref{eq:EOM} and obtain for the long-time limit the amplitude
\begin{align}\label{eq:response_X}
    \left|{x}(\omega)\right| = X(\omega) = \frac{F_0 /m}{\sqrt{\left(\omega_0^2 - \omega^2\right)^2 + \omega^2\Gamma^2}}
\end{align}
and the phase
\begin{align}\label{eq:response_phi}
    \phi(\omega) = \tan^{-1}\left(\frac{\omega\Gamma}{\omega_0^2-\omega^2}\right)\,.
\end{align}
In Fig.~\ref{fig:Fig2}, we show the result of such a measurement together with those of analytical expressions in Eqs.~\eqref{eq:response_X} and \eqref{eq:response_phi} for the fit values $f_0 = \omega_0/2\pi =\SI{28.52}{Hz}$ and $\Gamma =  \SI{0.209}{\hertz}$.

In the following, we will characterize and analyze our resonator in terms of the phase space quadratures $u$ and $v$, which fulfill $X^2 = u^2 + v^2$ and $\phi = \tan^{-1}(v/u)$~\cite{Eichler_Zilberberg_book}. These quadratures are the amplitudes of the two independent, fundamental solutions of the harmonic oscillator, $x = u\cos(\omega t) - v\sin(\omega t)$. Via the averaging method~\cite{Holmes1981161,Papariello_2016}, we can transform Eq.~\eqref{eq:EOM} into so-called slow-flow equations, which are coupled first-order differential equations of the form
\begin{align}
    &\dot{u} = -u\frac{\Gamma}{2} + v\frac{\omega^2-\omega_0^2}{2\omega} + F_0\frac{\sin(\theta)}{2m\omega}\,,\label{eq:slow-flow_1}\\
    &\dot{v} = -v\frac{\Gamma}{2} - u\frac{\omega^2-\omega_0^2}{2\omega} - F_0\frac{\cos(\theta)}{2m\omega}\,.\label{eq:slow-flow_2}
\end{align}
The slow-flow variables $u(t)$ and $v(t)$ correspond to the in-phase and out-of-phase channels of a lock-in amplifier and are therefore directly accessible in many experiments.

\section{Stochastic Response}
In order to observe stochastic physics, we replace the sinusoidal oscillating force by a fluctuating force of the form $F(t) = \xi(t)$, which has the properties
\begin{align}
	\left\langle \xi(t)\right\rangle &= 0\,,\\
	\left\langle \xi(t)\xi(t+\tau)\right\rangle &= \varsigma^2\delta(\tau)\,,\label{eq:xi}
\end{align}
where $\left\langle...\right\rangle$ denotes the average in the long-time limit, $\tau$ is a time delay, $\delta(...)$ is the delta function that is $1$ for $\tau = 0$ and $0$ otherwise, and $\varsigma^2$ has the meaning of a force noise power spectral density~\cite{Eichler_Zilberberg_book}.

The slow-flow Eqs.~\eqref{eq:slow-flow_1} and \eqref{eq:slow-flow_2} are correspondingly modified to read
\begin{align}
    &\dot{u} = -u\frac{\Gamma}{2} + v\frac{\omega^2-\omega_0^2}{2\omega} + \frac{\Xi_u}{m}\,,\label{eq:slow-flow_3}\\
    &\dot{v} = -v\frac{\Gamma}{2} - u\frac{\omega^2-\omega_0^2}{2\omega} +\frac{\Xi_v}{m}\,,\label{eq:slow-flow_4}
\end{align}
where the $\Xi_i$ are fluctuating force terms whose standard deviation can be calculated from an integration of $\xi$ over one oscillation period~\cite{Eichler_Zilberberg_book}. These fluctuating force terms are still white and random on timescales $\tau>1/f_0$, which is sufficient for all practical purposes.

\begin{figure}[t]
    \includegraphics[width=0.5\textwidth]{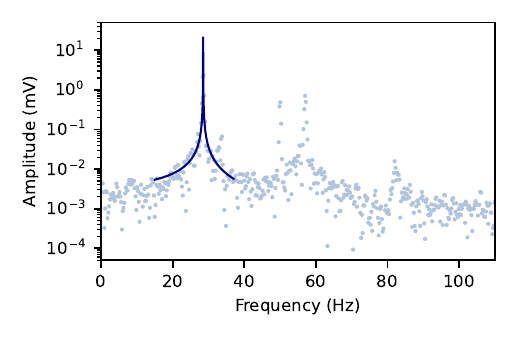}
    \caption{Frequency spectrum of the string driven by a fluctuating force with a standard deviation of \SI{500}{mV} over a bandwidth of \SI{30}{\mega\hertz}. We clearly see the resonance frequency $f_0$ at \SI{28.5}{Hz}, the \SI{50}{Hz} noise of the power grid, and a higher mode at roughly $2 f_0$. A dark blue line is shows the result of Eq.~\eqref{eq:response_X} with the same parameters as in Fig.~\ref{fig:Fig2}.}
    \label{fig:Fig3}
\end{figure}

The white force effectively drives the string at all frequencies simultaneously, producing the oscillation spectrum shown in Fig.~\ref{fig:Fig3}. In the frequency range between $0$ and \SI{100}{\hertz}, we observe four peaks. The mode at \SI{50}{\hertz} is assigned to purely electrical noise induced by the local power grid, while a peak below \SI{60}{\hertz} may belong to the second vibrational mode of the string. The lowest mode at \SI{28.5}{\hertz} is our vibration mode of interest. Using the known value for $f_0$ and $\Gamma$, we can reproduce its spectral signature with Eq.~\eqref{eq:response_X}, using only the effective amplitude $\varsigma/m$ as a free parameter.

\begin{figure}[t]
    \includegraphics[width=0.5\textwidth]{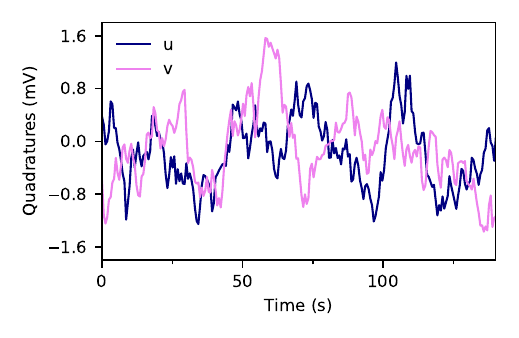}
    \caption{The two quadrature signals in response to a fluctuating force term, measured as a function of time with a lock-in amplifier whose local oscillator matches $f_0$.}
    \label{fig:Fig4}
\end{figure}

The time response of the lowest oscillator mode to a stochastic force is shown in Fig.~\ref{fig:Fig4}. In this measurement, the quadratures $u$ and $v$ are directly measured with a lock-in amplifier with a local oscillator (a reference clock) at \SI{28.5}{Hz}. Selecting a bandpass filter of \SI{4.5}{Hz} ensures that signals from higher modes are rejected, even though they are driven by the white noise as well. We find that the resonator no longer rings up to a fixed solution as in the deterministic case discussed earlier. Instead, the quadratures slowly fluctuate in time around the mean values $\left\langle u\right\rangle = \left\langle v\right\rangle = 0$ with a Gaussian profile.

\begin{figure}[t]
    \includegraphics[width=0.5\textwidth]{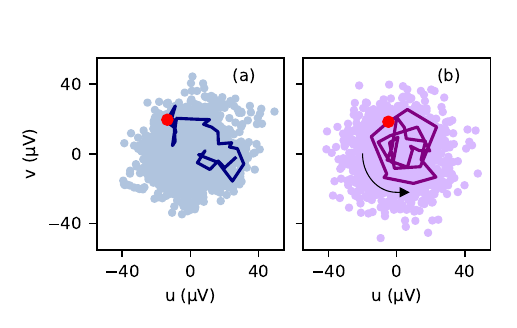}
    \caption{Phase space representation of the string oscillations driven by fluctuating force terms. The quadratures $u$ and $v$ are measured with in a frame rotating a the local oscillator frequency of the lock-in, set as (a)~\SI{28.5}{Hz} and (b)~\SI{28.4}{Hz} (b). The starting point of the measurement is marked as a red dot, and the first 30 points are connected with a line to demonstrate a random walk for $\omega = \omega_0$ and counter-clockwise rotation for $\omega < \omega_0$.}
    \label{fig:Fig5}
\end{figure}

To study the dynamics of the stochastically driven oscillator in more detail, we plot $v$ as a function of $u$ in Fig.~\ref{fig:Fig5}. This representation is known as ``phase space'' and plays an important role in the visualization of classical and quantum oscillation states. In our classical case, the system can be located at a single point in phase space at any given time, and a stable solution appears as a single fixed point. For instance, setting $\dot{u} = \dot{v} = 0$ and $\omega = \omega_0$ in Eqs.~\eqref{eq:slow-flow_1} and \eqref{eq:slow-flow_2} results in two simple equations $u \propto F_0 \sin(\theta)$ and $v \propto -F_0 \cos(\theta)$.

We plot the stochastic data similar to that in Fig.~\ref{fig:Fig4} in phase space, see Fig.~\ref{fig:Fig5}(a). The system explores phase space in a random walk over time, leading to Gaussian probability distributions in both coordinates. As we selected $\omega = \omega_0$ for this measurement, the terms $\propto \omega^2-\omega_0^2$ become zero in Eqs.~\eqref{eq:slow-flow_3} and \eqref{eq:slow-flow_4} and the random trajectories in both coordinates are decoupled. Setting $\dot{u} = \dot{v} = 0$, the equations can be individually simplified to $u \propto \Xi_u$ and $v \propto \Xi_v$. The trajectory of the system over time follows a random walk in both $u$ and $v$ independently.

Our observations become more intricate when we introduce a detuning of the measurement clock from the resonator's eigenfrequency, $\omega \neq \omega_0$. At first glance, the phase space representation in Fig.~\ref{fig:Fig5}(b) closely resembles the one in Fig.~\ref{fig:Fig5}(a). However, when we follow a short part of the trajectory, we find it to revolve around the origin with a defined orientation. This feature, which is absent in Fig.~\ref{fig:Fig5}(a), clearly stems from the terms $\propto \omega^2-\omega_0^2$ in Eqs.~\eqref{eq:slow-flow_3} and \eqref{eq:slow-flow_4}, which create a correlation between $u$ and $v$ with a time delay. We can directly see this delayed correlation in Eqs.~\eqref{eq:slow-flow_3} and \eqref{eq:slow-flow_4} by inserting $\Gamma = \Xi_{u,v}=0$ and $\omega^2-\omega_0^2 = (\omega-\omega_0)(\omega+\omega_0)\approx (\omega-\omega_0)2\omega$, which leads to the simple equations $\dot{u} = \Delta v$ and $\dot{v} = \Delta u$ with $\Delta \equiv \omega-\omega_0$. When initialized at $u_i = 1$ and $v_i = 0$, this system rotates around the origin with an angular frequency $\Delta = (\omega-\omega_0)$, such that a correlation between $u$ and $v$ appears after a quarter period (when $u=0$ and $v= \pm 1$). The sign of the correlation at this delay depends on the sign of $\Delta$, i.e., whether the rotation proceeds in clockwise or anti-clockwise direction.

\begin{figure}[t]
    \includegraphics[width=0.5\textwidth]{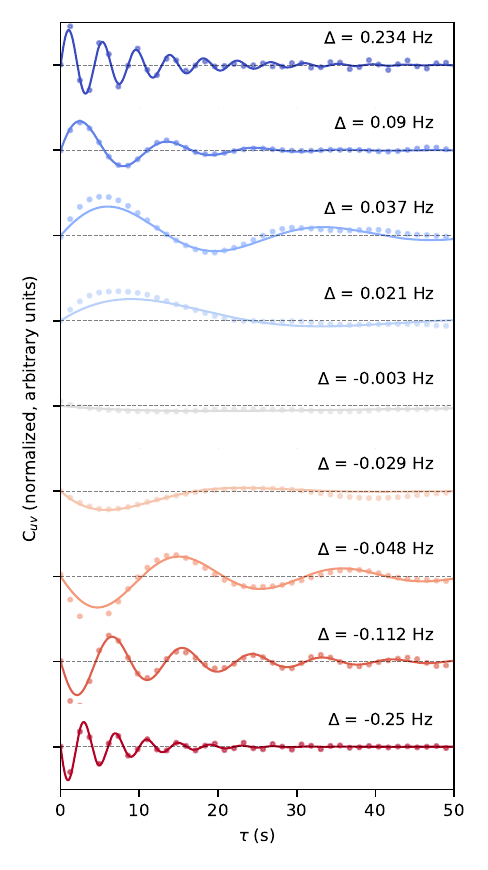}
    \caption{Time-dependent covariance $Cxx$ measured as a function of the time shift $\tau$ for different detuning $\Delta$. The data sets are offset vertically for better visibility, with dashed horizontal lines indicating $Cxx=0$ in all cases. Solid lines correspond to the function $e^{-\Gamma\tau / 2}\sin(\Delta\tau)$ with $\Delta$ fitted for each curve individually.}
    \label{fig:Fig6}
\end{figure}

To quantitatively analyze the effect of a detuned rotating frame in our data, we employ the delay-dependent covariance
\begin{align}\label{eq:tau_covariance}
    C_{uv}(n) = \frac{1}{N-n}\sum_{i=1}^{N-n}\left[ u(i)-\left\langle u\right\rangle\right]\left[v(i+n)-\left\langle v\right\rangle\right]
\end{align}
where $N$ is the total number of data points in a measurement, $i$ runs over all measurement points that are separated in time by the increment $\Delta_t$, and $\tau = n\Delta_t$ is a time delay. Equation~\eqref{eq:tau_covariance} is a covariance between $u$ and $v$ with a variable time delay $\tau$. We expect to see maxima in $C_{uv}$ whenever $u$ and $v$ values are correlated over a well-defined timescale $\tau$, indicating a rotation with a well-defined period.

The results of Eq.~\eqref{eq:tau_covariance} for measurements with different $\Delta$ are shown in Fig.~\ref{fig:Fig6}, together with a heuristic theory line corresponding to $C_{uv}=e^{-\Gamma\tau / 2}\sin(\Delta\tau)$, that is, an exponentially decaying oscillating correlation. Even though the full dependency of the covariance on $\tau$ is more complicated, this simple form reproduces our measurements qualitatively: detuning the measurement clock $\omega$ from the eigenfrequency $\omega_0$ leads to a periodic pattern in the measured time-dependent covariance $C_{uv}$, which decays under the influence of the dissipation term $\Gamma$. The detuning $\Delta$ therefore acts like a adjustable oscillation frequency in the rotating frame by generating correlations between $u$ and $v$. For a given value of $\Gamma$, we observe the rotating-frame analogue to an underdamped-to-overdamped transition between $\Delta > \Gamma$ (red and blue traces in Fig.~\ref{fig:Fig6}) and $\Delta \ll \Gamma$ (light orange to light blue traces). We have thus identified deterministic correlations in a system driven purely by fluctuating force terms.

\section{Summary and outlook}
In this work, we show how a simple and affordable guitar string setup can be used to demonstrate intricate physical phenomena. We start from the driven and damped harmonic oscillator, the most widely used model in physics, to describe the string oscillation to an external force. Building on this model, we introduce stochastic driving terms and observe the string's spectral response. We also study the stochastic dynamics of the string as a function of time and in phase space, which allows us to appreciate the notion of a Gaussian amplitude distribution in both quadratures and the influence of the selected rotating frame. Finally, we manipulate the rotating frame rate to obtain a rotating-frame analogue to an underdamped-to-overdamped transition in the quadrature correlations.

In our opinion, such a demonstration has high didactic value, as it provides hands-on experience with the driven and damped harmonic oscillator, deterministic versus stochastic physics, temporal versus spectral information, the rotating frame, and phase space representation. These topics are relevant for many fields of modern physics, such as quantum optics, nanomechanics, radio-frequency and microwave electronics, and gravitation wave experiments. From a teaching perspective, introducing these concepts with an experiment is a valuable supplement to traditional lectures.  In particular, the intuitive mechanics and macroscopic scale of the string resonator is beneficial to illustrate complex concepts in an accessible way.

Looking forward, we propose that the same string setup is also suitable to instruct students in nonlinear and parametric dynamics, especially the Duffing resonator, as well as parametric driving and squeezing~\cite{Leuch_2016}. These complex topics are often neglected in the experimental education of undergraduate students for lack of an affordable and easily maintainable experimental setup. Our guitar string offers a solution for this problem, and can serve as a catalyst to introduce students to a range of fascinating experimental techniques at an early stage in their curriculum.

\textbf{List of student tasks:}

1. Apply an oscillating drive voltage to the coil and measure the response voltage from the piezo with an oscilloscope. Look at the frequencies within the response by Fourier transforming it.

2. Change the frequency $f$ of the drive voltage and observe the change in the amplitude when this drive frequency is close to $f_0$.

3. Measure both the drive voltage and the response voltage at the same time with the oscilloscope. How can you measure the phase in this way?

4. Measure the amplitude and the phase with a lock-in amplifier and plot it as a function of the driving frequency $f$ (a free software lock-in module is available \href{https://pyrpl.readthedocs.io/en/latest/}{here}). Add matching lines corresponding to Eq.~\eqref{eq:response_X} and \eqref{eq:response_phi} and extract $\omega_0 = 2\pi f_0$ and $\Gamma$ in this way.

5. Replace the oscillating driving voltage with a white noise voltage and check that your noise is indeed white and has a Gaussian distribution in voltage. Measure the response of the resonator as a frequency spectrum. Fit Eq.~\eqref{eq:response_X} to your spectrum and extract $\omega_0$ and $\Gamma$. Compare the spectrum of the resonator response to that of the white noise itself.

6. Measure $u$ and $v$ with a lock-in amplifier for different local oscillator frequencies, that is, detunings $\Delta$. Plot the results in a rotating phase space as in Fig.~\ref{fig:Fig5} and determine the histograms in both quadratures. Pay attention to the sampling rate relative to $1/\Gamma$. What is an ideal sampling rate, and why? Does $\Delta$ impact the histograms?

7. Use Eq.~\eqref{eq:tau_covariance} to obtain a delay-dependent covariance for every measurement and extract the true $\Delta$ in each case. What is the origin of the deterministic order you find here?

\vspace{5mm}
\textit{The authors have no conflicts to disclose.}
\vspace{5mm}

Acknowledgements: the authors thank Oded Zilberberg for useful discussions. Oliver Schwager and the mechanical workshop at the Physics Department of ETH Zurich contributed to the development of the setup.

\providecommand{\noopsort}[1]{}\providecommand{\singleletter}[1]{#1}%
%


\end{document}